\documentstyle[12pt]{article}
\input epsf
\catcode`\@=11
\def\marginnote#1{}
\def\ifmath#1{\relax\ifmmode #1\else $#1$\fi}

\def\bold#1{\setbox0=\hbox{$#1$}%
     \kern-.025em\copy0\kern-\wd0
     \kern.05em\copy0\kern-\wd0
     \kern-.025em\raise.0433em\box0 }

\def\GENITEM#1;#2{\par\vskip6pt \hangafter=0 \hangindent=#1
   \Textindent{$ #2$ }\ignorespaces}

\newcount\hour
\newcount\minute
\newtoks\amorpm
\hour=\time\divide\hour by60
\minute=\time{\multiply\hour by60 \global\advance\minute by-
\hour}
\edef\standardtime{{\ifnum\hour<12 \global\amorpm={am}%
    \else\global\amorpm={pm}\advance\hour by-12 \fi
    \ifnum\hour=0 \hour=12 \fi
    \number\hour:\ifnum\minute<100\fi\number\minute\the\amorpm}}
\edef\militarytime{\number\hour:\ifnum\minute<100\fi\number\minute}
\def\draftlabel#1{{\@bsphack\if@filesw {\let\thepage\relax
  \xdef\@gtempa{\write\@auxout{\string
    \newlabel{#1}{{\@currentlabel}{\thepage}}}}}\@gtempa
    \if@nobreak \ifvmode\nobreak\fi\fi\fi\@esphack}
     \gdef\@eqnlabel{#1}}
\def\@eqnlabel{}
\def\@vacuum{}
\def\draftmarginnote#1{\marginpar{\raggedright\scriptsize\tt#1}}
\def\draft{\oddsidemargin -.5truein
        \def\@oddfoot{\sl preliminary draft \hfil
        \rm\thepage\hfil\sl\today\quad\militarytime}
        \let\@evenfoot\@oddfoot \overfullrule 3pt
        \let\label=\draftlabel
        \let\marginnote=\draftmarginnote

\def\@eqnnum{(\theequation)\rlap{\kern\marginparsep\tt\@eqnlabel}%
\global\let\@eqnlabel\@vacuum}  }
\def\preprint{\twocolumn\sloppy\flushbottom\parindent 1em
        \leftmargini 2em\leftmarginv .5em\leftmarginvi .5em
        \oddsidemargin -.5in    \evensidemargin -.5in
        \columnsep 15mm \footheight 0pt
        \textwidth 250mmin      \topmargin  -.4in
        \headheight 12pt \topskip .4in
        \textheight 175mm
        \footskip 0pt

\def\@oddhead{\thepage\hfil\addtocounter{page}{1}\thepage}
        \let\@evenhead\@oddhead \def\@oddfoot{} \def\@evenfoot{}
}
\def\titlepage{\@restonecolfalse\if@twocolumn\@restonecoltrue\o
necolumn
     \else \newpage \fi \thispagestyle{empty}\c@page\z@
        \def\thefootnote{\fnsymbol{footnote}} }
\def\endtitlepage{\if@restonecol\twocolumn \else  \fi
        \def\thefootnote{\arabic{footnote}}
        \setcounter{footnote}{0}}  
\catcode`@=12
\relax
\def\be{\begin{equation}}
\def\ee{\end{equation}}
\def\bea{\begin{eqnarray}}
\def\eea{\end{eqnarray}}
\def\simlt{\stackrel{<}{{}_\sim}}
\def\simgt{\stackrel{>}{{}_\sim}}

\def\mst11{m_{\;\widetilde{t}_{1}}}

\def\mbino{m_{\,\widetilde{B}}}
\def\mgluino{m_{\,\widetilde{g}}}
\def\msbL{m_{\,\widetilde{b}_L}}
\def\msbR{m_{\,\widetilde{b}_R}}
\def\mstauL{m_{\;\widetilde{\tau}_L}}
\def\mstauR{m_{\;\widetilde{\tau}_R}}

\def\sbm{m_{\;\widetilde{b}}}
\def\staum{m_{\;\widetilde{\tau}}}
\def\stm{m_{\;\widetilde{t}}}
\def\mst22{m_{\;\widetilde{t}_{2}}}
\def\mst12{m_{\;\widetilde{t}_{1,2}}}

\def\msb11{m_{\;\widetilde{b}_{1}}}
\def\msb22{m_{\;\widetilde{b}_{2}}}
\def\msb12{m_{\;\widetilde{b}_{1,2}}}
\def\msbl{m_{\;\widetilde{b}_L}}
\def\msbr{m_{\;\widetilde{b}_R}}

\def\mwidetilde2{\widetilde{m}^{2}}

\relax
\newcommand{\newc}{\newcommand}

\newc{\bit}{\begin{itemize}}
\newc{\eit}{\end{itemize}}
\newc{\ben}{\begin{enumerate}}
\newc{\een}{\end{enumerate}}

\newc{\gsim}{\lower.7ex\hbox{$\;\stackrel{\textstyle>}{\sim}\;$}}
\newc{\lsim}{\lower.7ex\hbox{$\;\stackrel{\textstyle<}{\sim}\;$}}
\newc{\thw}{\theta_W}
\newc{\ra}{\rightarrow}
\newc{\VEV}[1]{\langle #1 \rangle}
\newc{\hc}{{\it h.c.}}
\newc{\ie}{{\it i.e.}}
\newc{\etal}{{\it et al.}}
\newc{\eg}{{\it e.g.}}
\newc{\etc}{{\it etc.}}
\newc{\vrot}{v_{\rm rot}(r)}
\newc{\rhocrit}{\rho_{crit}}
\newc{\rhochi}{\rho_{\chi}}
\newc{\mpc}{~{\rm Mpc}}
\newc{\ev}{~{\rm eV}}     \newc{\kev}{~{\rm keV}}
\newc{\mev}{~{\rm MeV}}   \newc{\gev}{~{\rm GeV}}
\newc{\tev}{~{\rm TeV}}
\newc{\abund}{\Omega h^2_0}
\newc{\omegachi}{\Omega_\chi}
\newc{\abundchi}{\Omega_\chi h^2_0}
\newc{\sigmaann}{\sigma_{\rm ann}}
\newc{\vrel}{v_{\rm rel}}

\newc{\mtop}{m_{t}}
\newc{\mchi}{m_{\chi}}
\newc{\tanb}{\tan{\beta}}

\newc{\rgut}{r_{GUT}}
\newc{\hone}{H_1}
\newc{\htwo}{H_2}
\newc{\vev}{{\it v.e.v.}}
\newc{\vone}{v_1}   \newc{\vtwo}{v_2}
\begin{document}
\topmargin-1.cm
%
\begin{titlepage}
\vspace*{-64pt}
\begin{flushright}
{CERN-TH/98-130\\
OUTP-98-30-P\\
hep-ph/9804365\\
April 1998\\}
\end{flushright}

\vskip .7cm

\begin{center}
{\Large \bf  The Supersymmetric Flavour and CP Problems
\vskip 0.2cm From a Cosmological Perspective}

\vskip 1cm

{\bf Tony Gherghetta$^{a,}$\footnote{E-mail:
     {\tt  tony.gherghetta@cern.ch}}},
{\bf Antonio Riotto$^{a,}$\footnote{E-mail:
     {\tt riotto@nxth04.cern.ch}}$^,$\footnote{
     On leave of absence from the Department of Theoretical Physics,
     Oxford University, U.K. }}
and {\bf Leszek Roszkowski$^{b,}$\footnote{E-mail:
{\tt lr@virgo.lancs.ac.uk}}}

\vskip.3in
{\it 
$^a$Theory Division, CERN, CH-1211 Geneva 23, Switzerland\\
\vspace{3pt}
$^b$Department of Physics, Lancaster University, 
Lancaster LA1 4YB, U.K.\\ 
}

\end{center}

\vskip .5cm

\baselineskip=20pt

\begin{quote}

The supersymmetric flavour and CP problems can be avoided if the first 
two generations of sfermions are heavier than a few TeV and approximately 
degenerate in mass. However using flavour and CP-violating constraints 
on the third sfermion generation, together with the decoupling of the 
first two generations, can dramatically affect cosmological predictions 
such as the relic abundance of stable particles. In particular, we show 
that if the lightest supersymmetric particle is essentially bino-like 
then requiring that all flavour changing neutral current and CP-violating 
processes are adequately suppressed, imposes severe limits on the bino 
mass, where typically $\mbino\simgt (200-300) \gev$. This leads to 
difficulties for models implementing the scenario of heavy sfermion masses.

\end{quote}
\end{titlepage}

\setcounter{footnote}{0}
\setcounter{page}{0}
\newpage

\baselineskip=20pt

{\bf 1.} Supersymmetry (SUSY) is usually invoked  to solve many of the 
puzzles of the Standard Model such as the stability of the weak scale 
under radiative corrections. Furthermore, local
supersymmetry provides a promising way to include gravity within the
framework of unified theories of particle physics. For such reasons,  
supersymmetric extensions of the Standard Model have been the focus 
of intense theoretical activity in recent years~\cite{review}. 

Since experimental observations require supersymmetry to
be broken, it is essential to have a knowledge of
the nature and the scale of supersymmetry breaking in order to have a
complete understanding of the physical implications of these theories.
Unfortunately, at the moment we lack such an understanding and  
therefore it is important  to focus on the several experimental hints which 
might  be useful in  exploring the nature of supersymmetry breaking. 

The Minimal Supersymmetric Standard Model (MSSM) (or extensions of it)
are characterized by the presence of new degrees of freedom, the
scalar partners of the fermions (sfermions), which carry flavour
number and therefore can generate potentially large contributions to
the Flavour Changing Neutral Currents (FCNC's)~\cite{reviewfcnc}. 
Moreover, new CP-violating parameters may appear in
the low energy effective theory where SUSY is softly broken~\cite{reviewcp}.  
The requirement of consistency with the experimental
data imposes strong constraints on the physics of flavour and CP
violation in SUSY theories and has a profound impact on supersymmetric
model building. Although the flavour changing elements in the sfermion
mass matrices as well as the CP violating phases are free parameters
in the MSSM, ultimately their values have to be obtained from a theory
of soft supersymmetry breaking and fermion mass generation. Therefore,
experimental constraints provide us with useful suggestions towards
such a theory.

There are broad classes of solutions which solve the supersymmetric
flavour problem and the supersymmetric CP problem. The first
possibility is that for some deep theoretical reasons the pattern of
the sfermion mass matrices at the weak scale is very special: they are
either very close to the unity matrix in flavour space (flavour
universality)~\cite{dg} or they have a structure, but they are
diagonal in the basis set by the quark mass matrix (alignment)~\cite{ns}. 
Under these special conditions, the FCNC effects are tiny
and the CP violating phases at the weak scale are either highly
suppressed or efficiently screened~\cite{reviewcp}. Furthermore, if
high degeneracy of the first two sfermion generations occurs, their
masses are bounded from below only by the present direct searches.

The second and, {\it a priori}, the most straightforward possibility
occurs when the masses of the first and second generation of sfermions
are larger than a few TeV~\cite{dine,pom} and much larger than the
masses of sfermions of the third generation. In principle, this inverse 
hierarchy (compared to fermion masses) could be a consequence of the 
supersymmetry breaking pattern at the Planck or string scale 
\cite{brignole}. Other possibilities include integrating out heavy states 
which give rise to extra contributions to the soft mass terms of light 
particles \cite{heavy}. 
Notice, though, that the contribution to $\epsilon_K$ from the first two 
sfermion generations is generically still too large for CP violating phases 
${\sim\cal O}(1)$. However, this scenario becomes tenable when further 
approximate degeneracy in the mass spectrum of the first two generations of 
squarks is present, such as in models with non-abelian horizontal symmetries. 
Explicit realizations of this possibility are presented 
in~\cite{pom,u1,savoy}. In this way, the suppression of
FCNC effects in the MSSM is achieved 
and the SUSY contributions
to CP violating observables are small even for CP violating phases of
order unity.  Note also that having the
first and second generation of sfermions heavy does not necessarily
lead to naturalness problems, since the first two generations are
almost decoupled from the Higgs sector and, in the absence of
universality, the naturalness upper limits on supersymmetric particle
masses increase somewhat compared to the case when universality is
assumed~\cite{gian}.  Still, even without universality,
the charginos and
neutralinos are likely to be accessible at LEP2.\footnote{In theories
where the soft SUSY breaking parameters are generated at a high scale,
large masses for the sfermions of the first and second generation may
drive the scalar top mass squared to negative values at the weak scale
because of the two-loop renormalization group
evolution~\cite{murayama}. This, in turn, puts a strong lower bound on the
value of the scalar top mass squared at the high scale.}

However, models with the first two squark generations heavy may predict 
in the neutral $B$ system sizeable shifts from the Standard Model predictions
of CP asymmetries in the decays to final CP eigenstates~\cite{reviewcp}. 
In general, the supersymmetric contributions to FCNC's and to the CP
violating observables are expected to come from the third generation
of sfermions and they are typically close to the present experimental
bound. This means that, lower bounds on the masses of the third
generation sfermion may still be imposed from phenomenological
considerations in the scenario in which 
the first two squark generations are decoupled.

On the other hand, it is well known that by considering the
cosmological relic density of stable particles one can impose
significant bounds on the parameter space of a given model. In the
MSSM with R-parity conservation, the lightest supersymmetric particle
(LSP), is absolutely stable and its contribution to
the relic abundance $\Omega_{{\rm LSP}} h^2$ in the 
Universe~\cite{reviewlsp} may be
inconsistent with the bound $\Omega_{{\rm LSP}} h^2\sim 1$ 
implied by a (very conservative) lower bound of at least 10 billion years
on the age of the Universe. 
The relic abundance of the LSP is determined by its
annihilation cross section, which depends sensitively upon the masses
of the various particles mediating the annihilation processes. For
instance, in the case when the LSP is a bino-like neutralino, 
which we denote by $\chi$, 
large sfermion masses are typically inconsistent with the
cosmological bound $\Omega_\chi h^2\simlt 1$, unless the annihilation 
rate of the LSP into scalar and gauge bosons is efficient enough 
and/or near resonances.  It is
therefore reasonable to expect that combining the experimental bounds
on FCNC and CP violating phenomena with the bounds coming from
cosmological considerations will help us in significantly constraining
the parameter space of the MSSM.

In the present paper we will assume that the solution to the
supersymmetric flavour problem and the supersymmetric CP problem is
provided by the second class discussed above, namely by the scenario
where the first and second generation sfermion masses are in the
${\cal O}(10)$ TeV range and approximately degenerate in mass. We will 
show that when parameters are chosen
so that the LSP is predominantly a bino, the requirement 
$\Omega_\chi h^2\simlt 1$ often places a severe {\it lower } bound on the LSP
mass. This result may have rich implications for the class of
supersymmetric models which explain the suppression of the FCNC and CP
violating effects by decoupling the first two generations of
sfermions.
 
{\bf 2.} Before beginning the discussion of the cosmological 
bounds, let us briefly discuss what kind of limits one can infer from the 
FCNC and CP violating effects on the masses of the third sfermion generation.
We will generically assume that the third generation sfermions are
lighter than a TeV. While bounds on the stops are fairly weak, larger 
effects arise for the sbottom and stau. The stringest bound that one can 
obtain on the sbottom mass follows from the $\epsilon_K$
parameter of $K^0-{\bar K}^0$ mixing. In the limit that 
$\sbm\equiv \msbl\simeq\msbr$ the
bound resulting from the $\epsilon_K$ parameter is~\cite{hkt,masiero}
\begin{equation}
\label{eKbound}
\left({1\: {\rm TeV}\over\sbm}\right)^2 \left|V_{13}^Q V_{23}^Q V_{13}^D 
V_{23}^D\right| \sin\varphi_1\: f(\mgluino^2/\sbm^2)
\simlt 3.24\times 10^{-5} 
\end{equation}
where $V^{Q,D}$ are flavour mixing matrices (that define the rotations which 
diagonalise the quark mass matrix in the basis where 
$m_{\tilde{Q},\tilde{D}}^2$ are diagonal),
$\varphi_1={\rm Arg}(V_{13}^Q V_{23}^{Q*} V_{13}^D V_{23}^{D*})$ is a 
CP-violating phase and $f(x)\simeq 3840\:x f_6(x)-204 {\tilde f}_6(x)$. 
The functions $f_6(x)$ and ${\tilde f}_6(x)$ are defined as~\cite{hkt}
\begin{eqnarray}
  f_6(x)&=&{1\over 6(1-x)^5}(-18x\ln x-6\ln x-x^3+9x^2+9x-17)\\
  {\tilde f}_6(x)&=&{1\over 3(1-x)^5}(-6x^2\ln x-6x\ln x+x^3+9x^2-9x-1).
\end{eqnarray}
Notice that the bound (\ref{eKbound}) depends on the particular details of 
the flavour mixing. Since we are considering models that do not 
have any special mechanisms for the flavour and CP-structure, we will 
generically assume the CP-phase to be maximal with $\sin\varphi_1\sim 1$.
In order to understand how the magnitude of the off-diagonal matrix elements 
affects the bound we will compare our results with a CKM-like 
parameterisation of the mixing matrices of the form
\begin{eqnarray}
\label{mixingmatrix}
                V^{Q,D}=\left(\begin{array}{ccc}
                1 & \lambda &\lambda^3\\
                \lambda & 1 &\lambda^2\\
                \lambda^3 & \lambda^2& 1 
                \end{array}\right)
\end{eqnarray}
where $\lambda\sim 0.2$ is a Cabibbo-like angle. The bound (\ref{eKbound}) 
is very sensitive to the amount of mixing between the first two and third 
generations 
\footnote{Notice also that since we are assuming CP violating phases 
$\sim {\cal O}(1)$ the contribution to $\epsilon_K$ from the first two 
generations is much too large (even for large squark masses). Therefore
as previously mentioned in the Introduction, one requires some approximate 
universality to further suppress these contributions. In particular,
if the first and second generation squark masses are
degenerate up to ${\cal O}(\lambda^2)$, then these contributions will 
be sufficiently suppressed \cite{pom}.}.
For arbitrary parameterisations of the mixing matrix we will 
present our results by defining an average off-diagonal element 
${\overline V_1}\equiv\left|V_{13}^Q V_{23}^Q V_{13}^D V_{23}^D\right|^{1/4}
/(0.2)^{5/2}$, where ${\overline V_1}=1$ corresponds to the CKM 
parameterisation (\ref{mixingmatrix}). For the special limit 
$\sbm\simeq\mgluino$ the sbottom mass bound arising from (\ref{eKbound}) is
\begin{equation}
        \sbm \simgt 800\; \overline V_1^2 \; {\rm GeV}.
\end{equation}
\begin{figure}[ht]
\centerline{ \epsfxsize 3.25 truein \epsfbox {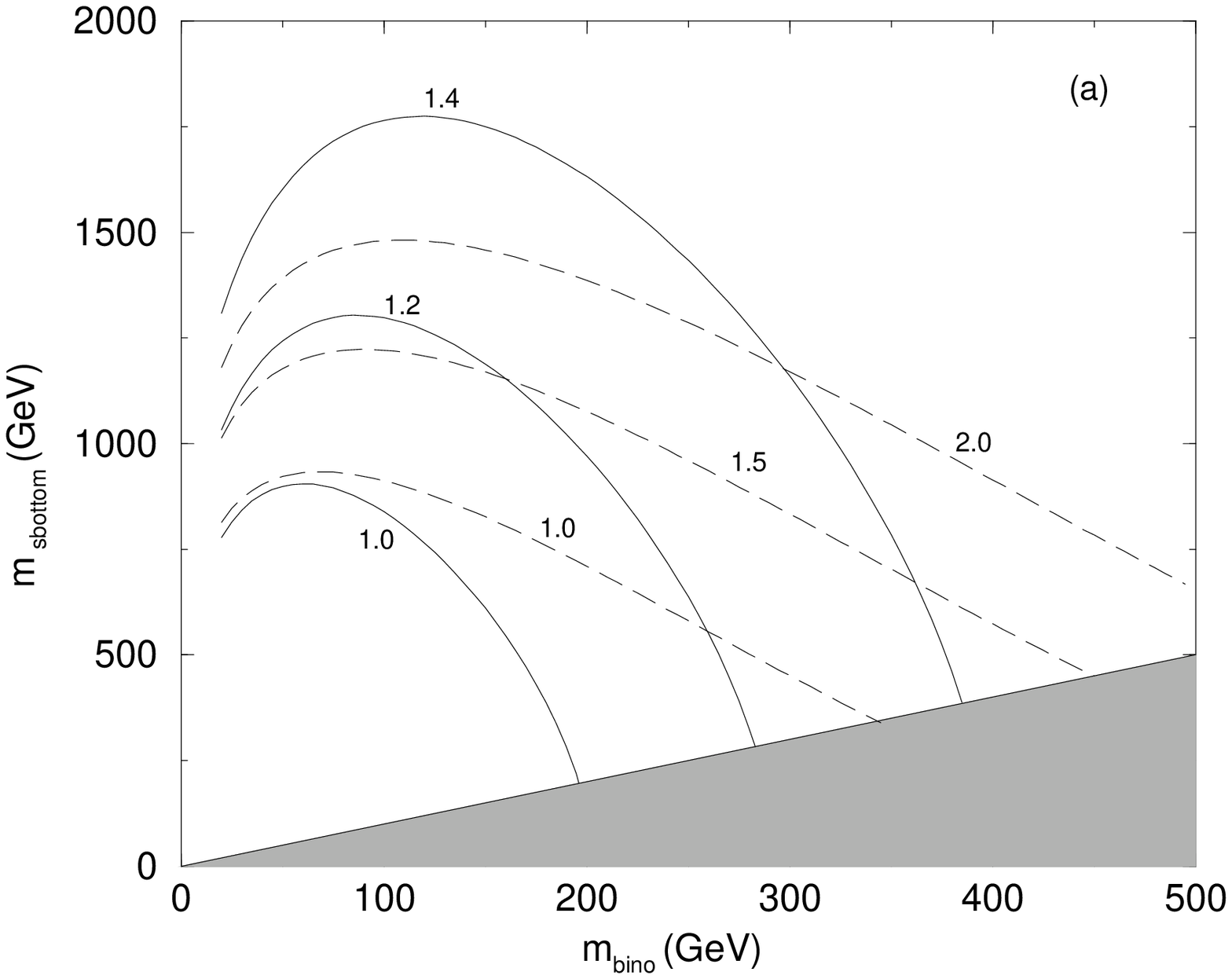}
             \epsfxsize 3.25 truein \epsfbox {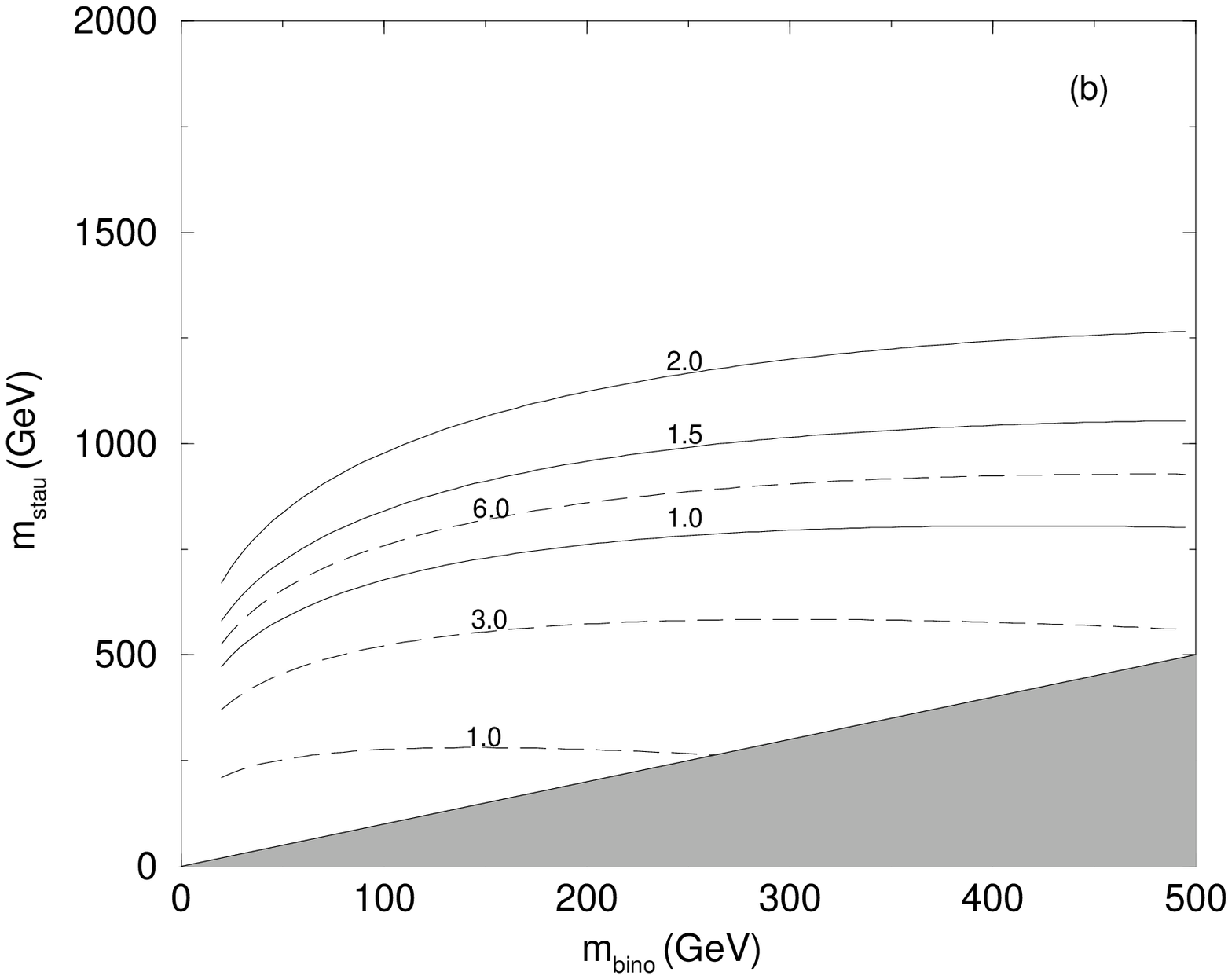}}
\caption{\it (a) Lower bounds on the sbottom mass for various contours of 
$\overline V_1$ (solid line) and $\overline V_2$ (dashed line) 
where $A'_b=10~{\rm TeV}$ and $\sin\varphi_{1,2}\sim 1$. (b)
Lower bounds on the stau mass for various contours of 
$\overline V_3$ (dashed line) and $\overline V_4$ (solid line) 
where $A'_\tau=1~{\rm TeV}$ and $\sin\varphi_4\sim 1$.}
\label{Vifigs}
\end{figure}
Clearly, the bound becomes weak when the amount of flavour mixing 
${\overline V_1}\rightarrow 0$. This is the case when there are special 
mechanisms operating such as universality or alignment. The general 
behaviour for arbitrary $\sbm$ and $\mgluino$ can be 
seen in Fig.~\ref{Vifigs}, where contours of the lower bound on the 
sbottom mass are shown for various values of ${\overline V_1}$. In the Figure,
$\sbm$ is plotted as a function of the bino mass $\mbino$ where
at the electroweak scale $\mgluino\simeq 7 \mbino$, which follows from 
our assumption of gaugino mass unification.

One can see that, for values of ${\overline V_1}\gsim 1$, 
the lower bounds on either the mass of the sbottom or the gluino is
quite significant, in the range of hundreds of GeV. Notice, however, that
the lower mass bound from $K^0-{\bar K}^0$ mixing disappears as 
the mass of the gluino or sbottom exchanged in the loop 
becomes very large. However, for large gluino 
mass a stronger lower bound can be obtained by considering the contribution 
of the sbottom left-right mixing to the down quark electric dipole moment
(EDM)~\cite{hkt,masiero}. This
contributes to the neutron EDM and gives rise to a bound
\begin{equation}
\label{nEDMbound}
{e \alpha_3 m_b\over 6\pi\sbm^4} \left|A_b^\prime\right|
\left|V_{13}^Q V_{13}^D \right|\sin\varphi_2\:\mgluino\:
\:g(\mgluino^2/\sbm^2)
\simlt 8.25\times 10^{-26} e {\rm \: cm}
\end{equation}
where $\varphi_2={\rm Arg}(V_{13}^Q V_{13}^D A_b^\prime)$,
$A_b^\prime=(A_b+\mu\tan\beta)$ and
\begin{equation}
  g(x)={1\over (1-x)^4}(2x^2\ln x+4x\ln x-5x^2+4x+1).
\end{equation}
The bound (\ref{nEDMbound}) is again sensitive to the flavour structure, 
which we will parameterise by ${\overline V}_2\equiv\left|V_{13}^Q V_{13}^D 
\right|^{1/2}/(0.2)^{3}$. However, unlike the bound arising from 
$K^0-{\bar K}^0$ mixing, the bound (\ref{nEDMbound}) is also sensitive to 
the amount of left-right mixing in the sbottom mass matrix. 
Thus besides the phases from the flavour 
mixing there is also a CP-phase from the left-right mixing. (The effect of 
a CP-phase in $A_b^\prime$ was previously considered in Ref.~\cite{fos} where 
it was shown that the limits on the LSP mass can be relaxed by a factor 
of 2-3.) Note that even in the absence of a CP-phase 
for $A'_b$, the bound (\ref{nEDMbound}) still applies provided there remain
nontrivial phases in the matrix elements $V_{13}^{Q,D}$.
Again, without considering any special mechanism for the 
CP-phases, we will assume that the overall CP-phase to be maximal 
($\sin\varphi_2\sim 1$). 
In the special limit $\sbm\simeq\mgluino$ we obtain
\begin{equation}
        \sbm \simgt 410\; \overline V_2^{2/3} \; 
        \left( {|A'_b|\over 1~{\rm TeV}} \right)^{1/3}\; {\rm GeV}.
\end{equation}
It is clear that strong constraints on the sbottom mass can only be 
obtained for large $A'_b$. This can be seen in Fig.~\ref{Vifigs} 
where $\overline V_2$ is plotted for $A'_b=10~{\rm TeV}$. 

Similar bounds can also be obtained for the stau and these follow from 
the flavour-violating process $\mu\rightarrow e\gamma$ and the 
electron EDM~\cite{hkt,masiero}. 
Again we will assume that $\staum\equiv\mstauL\simeq\mstauR$. 
The bound following from $\mu\rightarrow e\gamma$ is
\begin{equation}
\label{mue}
\left({100\: {\rm GeV}\over\staum}\right)^2 V_{13}^L V_{23}^E
{\mbino\: m_\tau\: \left|A_\tau^\prime\right| \over \staum^2 m_\mu}
\:g(\mbino^2/\staum^2)\simlt 5\times 10^{-4}
\end{equation}
where $A_\tau^\prime=(A_\tau+\mu\tan\beta)$ and $V^{L,E}$ are the slepton 
mixing matrices. Again for comparison we will assume a CKM-like 
parameterisation of the matrices $V^{L, E}$ which define the rotations 
that diagonalise the lepton mass matrix in the basis where the slepton 
mass matrices are diagonal. Thus ${\overline V}_3\equiv
(V_{13}^L V_{23}^E)^{1/2}/(0.2)^{5/2}$ defines an average off-diagonal 
matrix element normalised 
to a CKM-like parameterisation. Notice also that there is no CP-phase 
since the process is CP-invariant. The typical size of the bound on 
the stau mass can be obtained from considering the limit
$\staum\simeq\mbino$, where
\begin{equation}
        \staum \simgt 260\; \overline V_3^{2/3} \; 
        \left( {|A'_\tau|\over 1~{\rm TeV}} \right)^{1/3}\; {\rm GeV}.
\label{approxV3}
\end{equation}
Despite the insensitivity to 
CP-phases, the bounds arising from (\ref{mue}) only become strong for 
$|A'_\tau|\gg 1$ TeV or ${\overline V}_3\gg 1$, as can be ascertained 
from (\ref{approxV3}) and Fig.~\ref{Vifigs}. 

Much stronger constraints on the stau mass can be obtained from the 
electron EDM. The bound resulting from the electron EDM is~\cite{hkt,masiero}
\begin{equation}
\label{eEDMbound}
{e \alpha_1 m_\tau\over 2\pi\staum^4} \: \left| A_\tau^\prime\right| 
\left|V_{13}^L V_{13}^E \right| \sin\varphi_4\: \mbino\:
g(\mbino^2/\staum^2) \simlt 7\times 10^{-27} e {\rm \: cm}
\end{equation}
where $\varphi_4={\rm Arg}(V_{13}^L V_{13}^{E*} A_\tau^\prime)$.
Again assuming the CP-phase to be maximal ($\sin\varphi_4\sim 1$) and 
defining ${\overline V}_4 = \left|V_{13}^L V_{13}^E\right|^{1/2}/(0.2)^3$, 
lower bounds on the stau mass can be obtained for large left-right 
mixing $(A_\tau^\prime)$ in the stau mass matrix. In particular for the
limit $\staum\simeq\mbino$ we see that
\begin{equation}
        \staum \simgt 750\; \overline V_4^{2/3} \; 
        \left( {|A'_\tau|\over 1~{\rm TeV}} \right)^{1/3}\; {\rm GeV}.
\end{equation}
In Fig.~\ref{Vifigs}, 
the stau mass bound is shown for contours of $\overline V_4$ and 
$A'_\tau=1~{\rm TeV}$. Unlike $\overline V_3$, the bounds arising from
$\overline V_4$ are always much stronger for the same value of $|A'_\tau|$
and $\overline V_3\simeq \overline V_4$, using the current experimental
bounds.

In all the above bounds we have made the degenerate squark mass assumption of 
$\sbm\equiv\msbL\simeq \msbR$ at the electroweak scale, 
and similarly for the stau. If this assumption 
is relaxed then the bounds shown in the Figures are for the geometric mean 
$\sqrt{\msbL\msbR }$ up to factors of $\cal O$(1) which follow from 
generalising the functions $f(x)$ and $g(x)$. It is also clear that if there
are any special mechanisms operating in the flavour structure, such that
$\overline V_i\rightarrow 0$, then all FCNC and CP-violating
bounds disappear. However, we will be specifically interested in the
cosmological consequences of the case where the first two sfermion
generations are heavy and generically $\overline V_i \sim {\cal O}(1)$.

{\bf 3.} Let us now consider the cosmological implications 
on the bino mass from the stringent lower bounds on the mass of the
third generation sfermions resulting from the 
FCNC and CP-violating processes. We will be 
particularly interested in the cosmological relic abundance of the
LSP when it is a neutralino which is predominantly bino-like,
with only a small admixture of the wino and the higgsino in its 
composition. While in principle any
superpartner could be the LSP, in the MSSM the neutralino is usually
{\em assumed} to be the LSP for astrophysical reasons: it is a
weakly-interacting stable massive particle for which astrophysical
bounds are very weak and it can  serve as an excellent dark matter
candidate~\cite{ehnos} when it is mostly a bino~\cite{chiasdm}. (Note that
a higgsino-like neutralino with a sufficiently large $\Omega_\chi h^2$
and a reasonably small mass has now been basically excluded by LEP-II,
except for a small remaining region~\cite{higgsino:new}.)

A predominantly bino-like LSP 
corresponds to the case $|\mu|\gsim M_1$ where $M_1$ is
the soft-mass of the bino.  It is worth noting that a bino-like
neutralino naturally arises as the only neutral LSP as a result of
requiring radiative electroweak symmetry breaking (EWSB).  While this
has been shown to be true mainly in the case of universal soft
masses at the unification scale~\cite{kkrw}, there are good reasons to
believe that this will also remain valid in the case studied here.
This is because $M_1$ depends on sfermion masses only at two loops,
while the parameter $\mu^2$ is determined via the condition for EWSB
where the sfermion masses of the first two generations enter
only as a small correction~\cite{gian}, and
are not expected to significantly alter the resulting value of $\mu$
compared to the universal case.

In order for a bino-like neutralino to give $\Omega_\chi h^2\sim1$, 
at least some sfermion masses should normally not exceed 
a few hundred GeV~\cite{chiasdm}. In our numerical analysis we will
include all relevant final states of the
neutralino annihilation and all exchange channels for the general case
of any neutralino composition. However, in the nearly pure bino limit
the dominant annihilation channel is into final state
(ordinary) charged fermions via the (lightest) sfermion exchange
and the relic abundance is 
approximately given by $\Omega_\chi h^2\propto m^4_{\tilde f}/\mchi^2$
where $m_{\tilde f}$ is the sfermion mass.
Thus it is clear that for sufficiently large sfermion masses imposing 
the bound  $\Omega_\chi h^2<1$ will imply a {\em lower} 
bound on $\mchi$, unless other final-state channels can reduce the 
LSP relic abundance below one. In the pure bino limit, the annihilation 
cross-section into final states involving one or both gauge bosons 
vanishes.  The final states involving the pseudoscalar $A$ and either 
$h$ or $H$, may be able to reduce $\Omega_\chi h^2$
below one, but they are not kinematically allowed 
until $\mchi\gsim (m_A+m_h)/2$. This implies a rather large $\mchi$ 
if $A$ is heavy.

The neutralino relic density is also reduced in the vicinity of
resonances due to the exchange of the $Z$ and the Higgs bosons. Again,
while the pure bino does not couple to the gauge or Higgs bosons, the
small higgsino component in the nearly pure bino case allows the
resonances to play some r{\^ o}le in decreasing $\Omega_\chi h^2$. 
Of special importance is the exchange of the pseudoscalar $A$
whose coupling to the neutralino is proportional to $\tanb$ and
therefore can become significantly enhanced, especially for larger
values of $\tanb$.

\begin{figure}[ht]
\centerline{\epsfxsize 5.8 truein \epsfbox {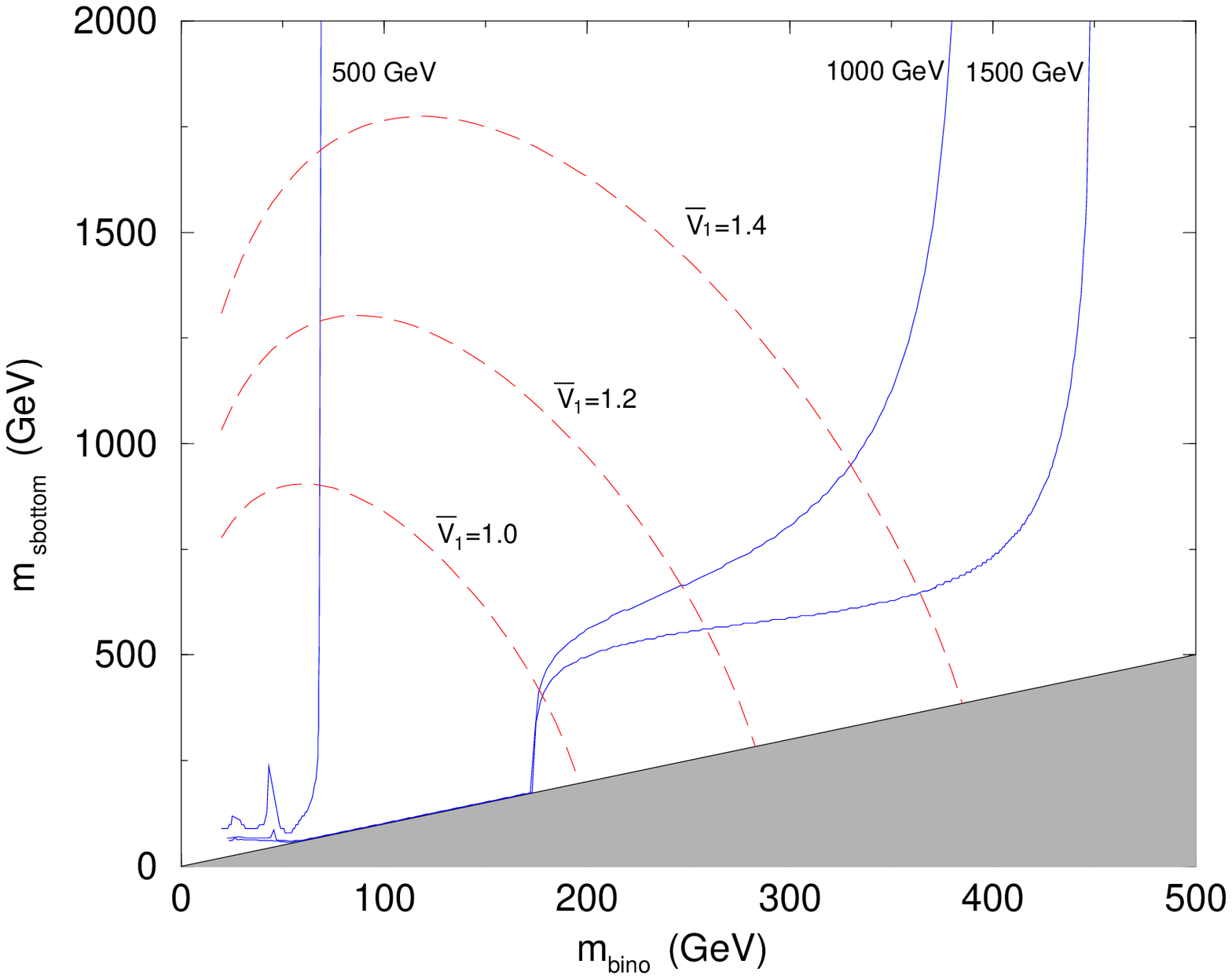}}
\caption{\it Bounds on the sbottom mass as a function of the bino mass. 
The $\overline V_1$ contours arise from the $\epsilon_K$ parameter 
of $K^0-\bar K^0$ mixing (regions below them are excluded). 
The cosmological contours 
$\Omega_\chi h^2=1$ are labelled by various values of the $\mu$
parameter. (Regions to the left and above them are excluded.)
In the Figure we have assumed $\stm=\sbm$, $\tan\beta=2$ and 
$m_A=\staum=1~{\rm TeV}$.}
\label{sbfig}
\end{figure}

\begin{figure}[ht]
\centerline{\epsfxsize 5.8 truein \epsfbox {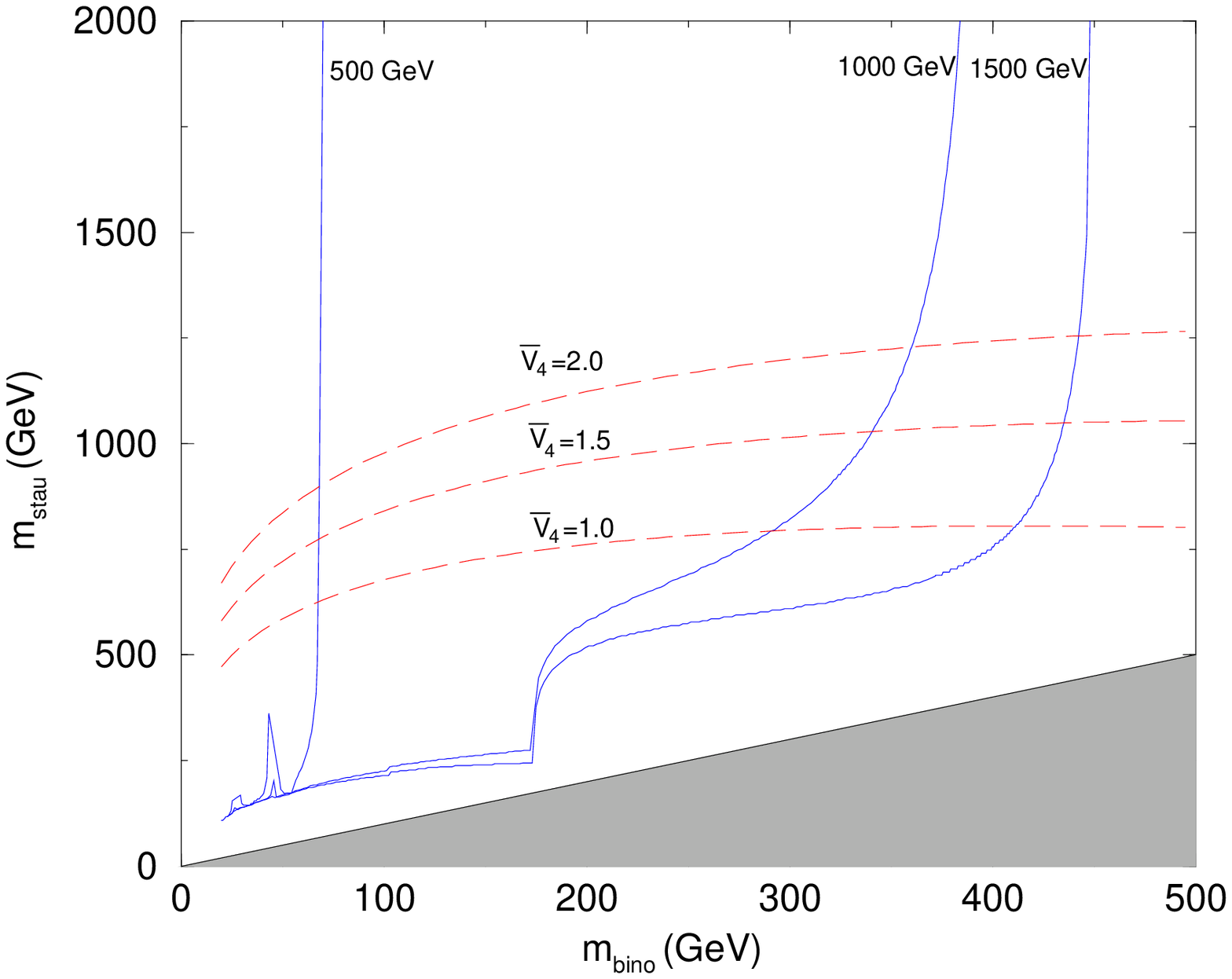}}
\caption{\it Bounds on the stau mass as a function of the bino mass. 
The $\overline V_4$ contours arise from the electron EDM  (regions
below them are excluded). The 
cosmological contours $\Omega_\chi h^2= 1$ are labelled by various 
values of the $\mu$ parameter. (Regions to the left and above them are
excluded.) 
In the Figure we have assumed $\stm=\staum, \tan\beta=2$ and $m_A=\sbm=
A'_\tau=1~{\rm TeV}$.}
\label{staufig}
\end{figure}

\begin{figure}[ht]
\centerline{\epsfxsize 5.8 truein \epsfbox {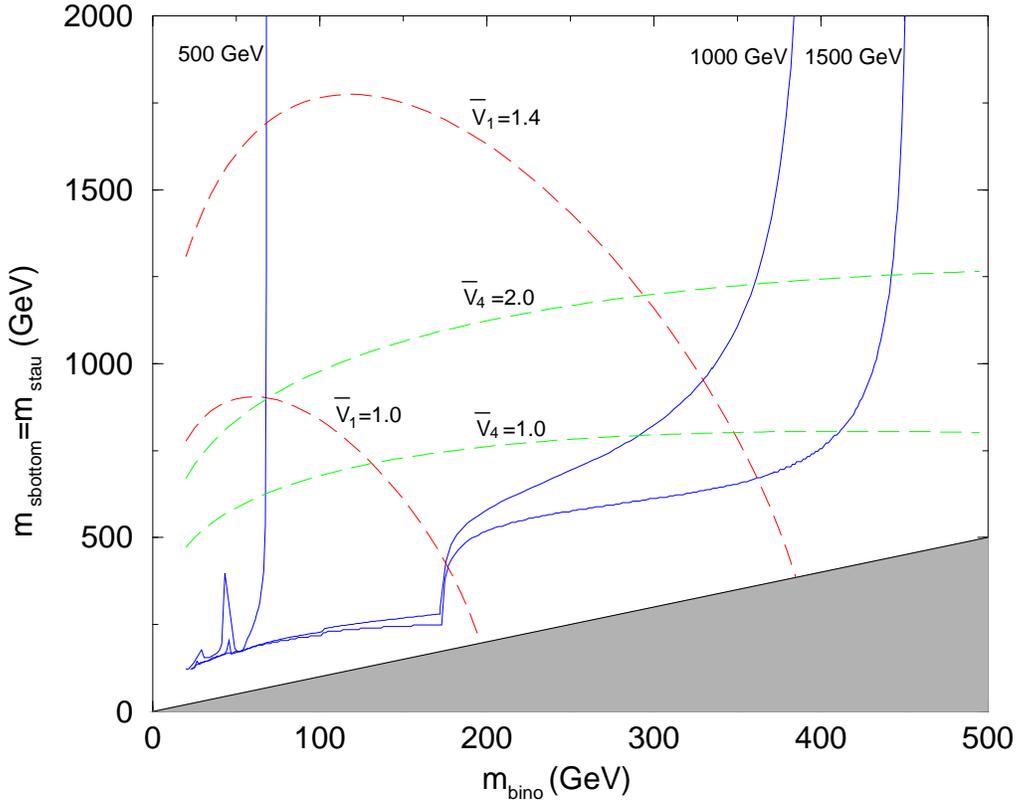}}
\caption{\it Bounds on the stau and sbottom 
mass as a function of the bino mass. The contours $\overline V_1$ 
are for the sbottom mass, while $\overline V_4$ constrains the stau mass. 
The cosmological contours $\Omega_\chi h^2= 1$ are labelled by 
various values of the $\mu$ parameter.
In the Figure we have assumed $\stm=\staum=\sbm, \tan\beta=2$ and $m_A=
A'_\tau=1~{\rm TeV}$.}
\label{sbstau1000fig}
\end{figure}

\begin{figure}[ht]
\centerline{\epsfxsize 5.8 truein \epsfbox {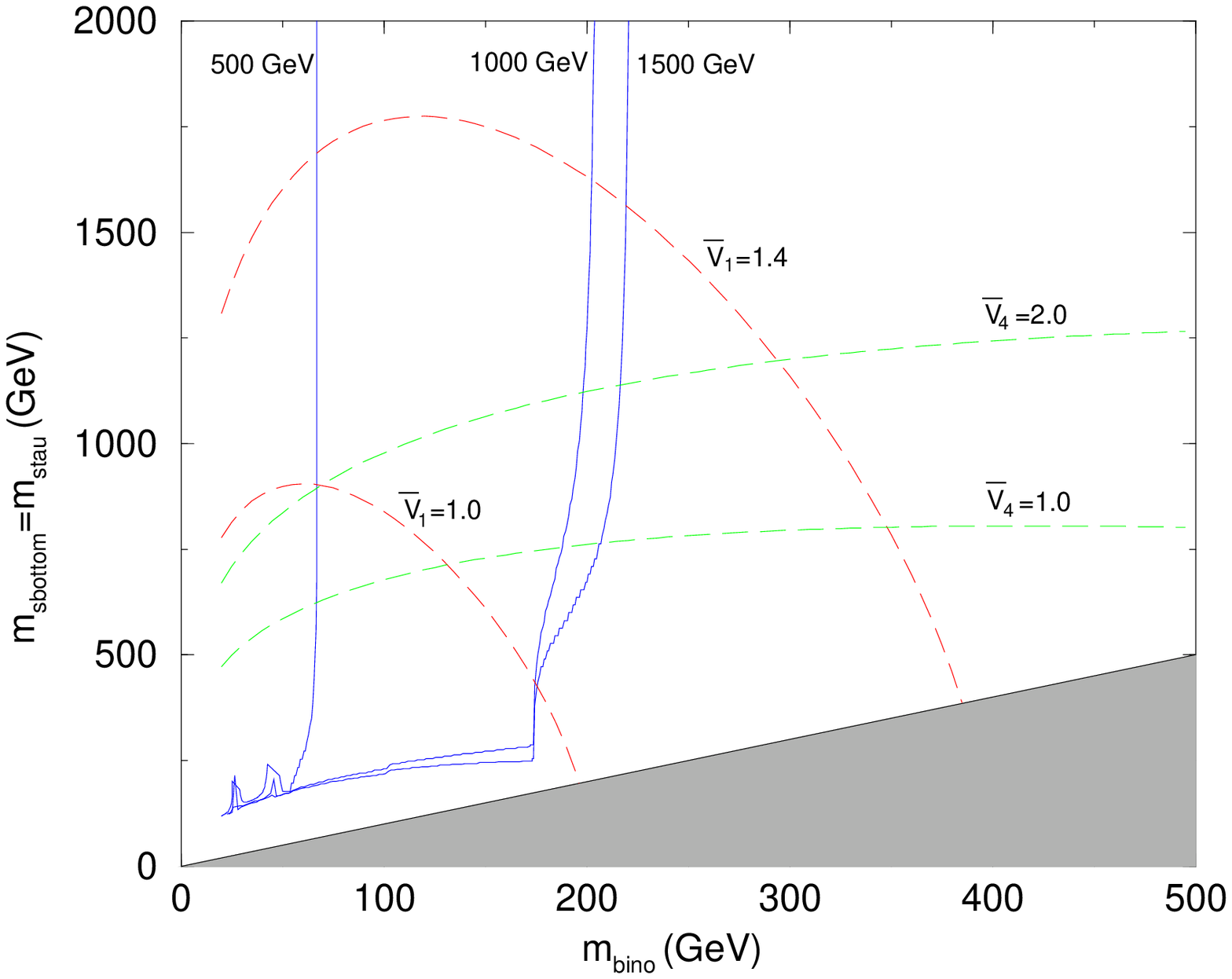}}
\caption{\it Same as Fig.~\ref{sbstau1000fig} except that 
$m_A=500~{\rm GeV}$.}
\label{sbstau500fig}
\end{figure}

Let us now combine the stringent limits on the masses of the third 
sfermion generation arising from the suppression of the FCNC and 
CP-violating processes with the cosmological constraint 
$\Omega_\chi h^2\simlt 1$ for a predominantly bino-like LSP.
We will consider three representative cases: $\sbm=\stm$ with 
$\staum$ heavy in Fig.~\ref{sbfig}, $\staum=\stm$ with $\sbm$ heavy in
Fig.~\ref{staufig} and $\sbm=\staum=\stm$ in Figs.~\ref{sbstau1000fig}
and ~\ref{sbstau500fig}. In each case we have used the best possible
constraint arising from FCNC and CP-violating processes. For the
sbottom mass this corresponds to the $\epsilon_K$ parameter,
parameterised by contours of $\overline V_1$, while for the stau mass 
the electron EDM parameterised by $\overline V_4$ provides the most 
stringent constraint. The cosmological contour 
$\Omega_\chi h^2=1$ is shown for several choices of $\mu$. 
Thus regions above and to the left of the cosmological contour
are excluded. 

In each Figure we see that as $|\mu|$ decreases, the higgsino
component of the neutralino increases, and consequently the
two-boson (both gauge and Higgs) final states become important. This
is especially true for the $ZZ$ and $WW$ final states which open up
for relatively low $\mchi$ but decouple in the pure bino limit. Since
we focus on a nearly pure bino as the LSP, we do not consider values of
$|\mu|$ smaller than 500~GeV in order for the bino purity 
(defined as the square of the bino
component in the neutralino mass eigenvector) to remain above $97\%$.  
For large $m_A$ and $|\mu|$ of the order of
1\tev\ and for small bino masses, below $\mtop$, only the tau and bottom
final states are effectively open and $\Omega_\chi h^2$ quickly
increases with the mass of their scalar partners, thus either leaving no room
for $\mchi<\mtop$ in Fig.~\ref{sbfig}, or allowing only for a
relatively narrow strip below $m_t$ in Fig.~\ref{staufig}.
When $\mchi>\mtop$, the $t\bar t$ channel opens up and is enhanced by 
the factor $\left(\mtop/m_W\right)^2$ via the higgsino component of the 
LSP. As the third generation sfermion masses increase further, this channel 
also becomes less and less effective. Finally, for $\mchi$ approaching $m_A/2$ 
the wide pseudoscalar resonance starts dominating quickly reducing
$\Omega_\chi h^2$ well below one.

The combination of the exclusion curves from flavour and CP violating
processes and from $\Omega_\chi h^2<1$ gives therefore strong
{\em lower} limits on $\mchi$. The limits are particularly strong for
large values of $|\mu|$ and $m_A$. For example, in Fig.~\ref{sbfig}
we see that for $|\mu|\simgt 1000\gev$ the bino has to be heavier than roughly
$\mtop$ even for $\overline V_1=1$. This should be compared with the
indicative upper bounds $\mchi\lsim 65\gev$,
obtained by requiring no significant fine-tuning in the 
parameters of the MSSM~\cite{gian}. 
Actually, since the motivation for this scenario is to
allow for basically unconstrained entries in the mixing matrices, one
would expect $\overline V_1$ significantly larger than one, in which case
the lower limit on $\mchi$ would be further significantly
increased. 

A similar picture emerges when one considers the bounds on the stau mass
arising from the electron EDM. Since the bounds on $\staum$ 
from $\overline V_4$
are more stringent than $\overline V_1$ we obtain a stronger lower
limit on the bino mass. For the case plotted in Fig.~\ref{staufig}
we find $\mchi\simgt 300\gev$ for $\overline V_4=1$ and $|\mu|
\simgt 1000\gev$. Finally in Figs.~\ref{sbstau1000fig} 
and~\ref{sbstau500fig} the sbottom and stau are now both assumed to be light
and we need to simultaneously satisfy the constraints on the sbottom and
stau from the suppression of FCNC and CP-violating processes. In this case 
since $\overline V_4$ sets the best limit we again find that 
$\mchi\simgt 300\gev$ for $|\mu|\simgt 1000$ GeV and $m_A=1000$ GeV
as shown in Fig.~\ref{sbstau1000fig}.

There are a number of ways one can relax the bounds on $\mchi$. This
can be done by either decreasing $|\mu|$ (thus increasing the higgsino
component of the LSP) or by lowering $m_A$ as can be seen in 
Fig.~\ref{sbstau500fig}. In this case the lower limit on $\mchi$
reduces to $\sim 200\gev$ for $|\mu|\simgt 1000\gev$ 
and is fairly independent of the value of 
$\overline V_1$ and $\overline V_4$. Another possibility is to
increase $\tanb$ in which case the resonance effect around $m_A/2$
widens considerably. Finally, while the sbottom and stau are
constrained by flavour and CP constraints, there are no constraints
on the stop. One can therefore choose to make the stop lighter than the 
sbottom and the stau. This can still only allow for $\mchi$ above 
$\mtop$ which is already a very strong lower bound. On the other hand,
we have found that for $\mu<0$ the bounds 
are even more stringent.

{\bf 4.} We have shown that by combining the constraints
arising from the suppression of FCNC and CP-violating processes
with bounds on the cosmological relic abundance, the bino mass
can be severely restricted. This places severe limitations on models
in which the first two sfermion generations are heavy and almost degenerate 
in mass and the supersymmetric contributions to the FCNC's and 
CP violating observables mainly come from the third squark  generation.  

Such a mass 
spectrum has been argued to be the best from the phenomenological point 
of view \cite{f} and may be obtained if the three families belong to a 
${\bf 2}+ {\bf 1}$ representation of a horizontal symmetry group $G_H$. 
For example, a class of models based on the group $U(2)$ predicts  
very heavy first and second family scalars, the CKM parameterisation 
(\ref{mixingmatrix}) of the mixing matrices, {\it i.e.} 
$\overline V_i=1$, and CP-violating phases of order unity \cite{pom}. 
It has also been recently pointed out that $D$-term contributions from 
the anomalous $U(1)$ gauge group in string theory may 
naturally lead to such a mass spectrum for the sfermions. 
On the other hand, a generic problem of this class of models is the 
generation of sizeable gaugino masses. In this paper we have pointed 
out that having the first two generations of sfermions heavy and 
approximately degenerate requires driving the mass of the bino-like 
LSP to quite large values when considerations 
about the present cosmological abundance of the LSP are taken into account. 
This leads to serious difficulties for models implementing the scenario 
of heavy sfermion masses. 

Our constraints can be avoided in a number of ways. First, if 
there is a small amount of R-parity violation then the LSP can  
simply decay and therefore be eliminated. In this case other solutions to 
the dark matter problem need to be considered. Secondly, one can envisage 
models where the mixing between the third generation and the 
first two generations of sfermions is small. For example, in certain 
three generation string solutions, the anomalous $U(1)$ couples 
universally to all three families \cite{fp}, yielding squark degeneracy.  
It is also possible to increase the higgsino or wino content of the 
neutralino LSP, but as we have mentioned earlier this scenario may
not be very natural from the point of view of mass unification.

\section*{Acknowledgements}

Two of us, TG and AR, would like to thank the Department of Physics
at Lancaster University for kind hospitality where 
part of this work was done.



\vskip .5in
\begin{thebibliography}{99}

\bibitem{review} See for example,  
H.P. Nilles, Phys. Rep. {\bf 110}, 1 (1984); H. Haber
and G. Kane, Phys. Rep. {\bf 117}, 76 (1984);
S.P. Martin, hep-ph/9709356.

\bibitem{reviewfcnc} For a recent review, see, {\it e.g.} 
M. Misiak, S. Pokorski and
J. Rosiek, {\it Supersymmetry and FCNC Effects}, in the Review Volume
{\it Heavy Flavours II}, edited by A.J. Buras and M. Lidner, World
Scientific Publishing, Singapore, hep-ph/9703442.

\bibitem{reviewcp} For a review, see Y. Grossman, Y Nir and R. Rattazzi, 
{\it CP violation Beyond the Standard Model}, in the Review Volume 
{\it Heavy Flavours II}, edited by A.J. Buras and M. Lidner, World 
Scientific Publishing, Singapore, hep-ph/9701231. 

\bibitem{dg} S. Dimopoulos and H. Georgi, Nucl. Phys. {\bf B193}, 150 (1981).

\bibitem{ns} Y. Nir and N. Seiberg, Phys. Lett. {\bf B309}, 337 (1993).

\bibitem{brignole} A. Brignole, L. Ibanez and C. Munoz, Nucl. Phys. 
{\bf B422}, 125 (1994).

\bibitem{heavy} Y. Kawamura et al., Phys. Rev. {\bf D51}, 1337 (1995); 
A. Pomarol and S. Dimopoulos, Nucl. Phys. {\bf B453}, 83 (1995); 
R. Rattazzi, Phys. Lett. {\bf B375}, 181 (1996). 

\bibitem{dine} M. Dine, A. Kagan and R.G. Leigh, Phys. Rev. {\bf D48}, 4269 
(1993).

\bibitem{pom} A. Pomarol and D. Tommasini, Nucl. Phys. {\bf B466}, 3 (1996).

\bibitem{u1} A.G. Cohen, D.B. Kaplan and A.E. Nelson, Phys. Lett. {\bf B388}, 
588 (1996); P. Binetruy and E. Dudas, Phys. Lett. {\bf B389}, 503 (1996);  
G. Dvali and A. Pomarol, Phys. Rev. Lett. {\bf 77}, 3728 (1996);  
R. N. Mohapatra and A. Riotto, Phys. Rev. {\bf D55}, 1138 (1997); 
Phys. Rev. {\bf D55}, 4262 (1997); A.E. Nelson and D. Wright,
Phys. Rev. {\bf D56}, 1598 (1997); 
N. Arkani-Hamed, M. Dine, S.P. Martin, hep-ph/9803432. 

\bibitem{savoy} E. Dudas and C. Savoy, Phys. Lett. {\bf B369}, 255 (1996); 
E. Dudas, C. Grojean, S. Pokorski and C. Savoy, Nucl. Phys. {\bf B481}, 85 
(1996).

\bibitem{gian} S. Dimopoulos and G.F. Giudice, Phys. Lett. {\bf B357}, 573  
(1995).

\bibitem{murayama} N. Arkani-Hamed and H. Murayama, Phys. Rev. {\bf D56}, 
6733 (1997).  

\bibitem{reviewlsp} For a review, see G. Jungman, M. Kamionkowski 
and K. Griest, Phys. Rept. {\bf 267}, 195 (1996).  

\bibitem{hkt} J. Hagelin, S. Kelley and T. Tanaka, Nucl. Phys. {\bf B415}, 
293 (1994).

\bibitem{masiero} F. Gabbiani, E. Gabrielli, A. Masiero, L. Silvestrini,
Nucl.Phys. {\bf B477}, 321 (1996).

\bibitem{fos} T. Falk, K. Olive and M. Srednicki, Phys. Lett. {\bf B354},
99 (1995).

\bibitem{ehnos} J.~Ellis, J.S.~Hagelin, D.V.~Nanopoulos, K.A.~Olive,
and M.~Srednicki, Nucl. Phys. {\bf B238}, 453 (1984).

\bibitem{chiasdm} L.~Roszkowski, Phys. Lett. {\bf B262}, 59 (1991).

\bibitem{higgsino:new} 
J. Ellis, T. Falk, G. Ganis, K.A. Olive and M. Schmitt, hep-ph/9801445.

\bibitem{kkrw} R.~Roberts and L.~Roszkowski, Phys. Lett. {\bf B309}, 329 
(1993); G.~Kane, C.~Kolda, L.~Roszkowski, J.~Wells, Phys. Rev. {\bf D49}, 
6173 (1994). 

\bibitem{f} See A.G. Cohen, D.B. Kaplan and A.E. Nelson in \cite{u1}.

\bibitem{fp} A.E. Faraggi and  J.C. Pati, hep-ph/9712516. 


\end{thebibliography}
\end{document}